# Types for Tables: A Language Design Benchmark

Kuang-Chen Lu[a], Ben Greenman[a], and Shriram Krishnamurthi[a]

a  Brown University, Providence, RI, USA

**Abstract**

**Context**  Tables are ubiquitous formats for data. Therefore, techniques for writing correct programs over tables, and debugging incorrect ones, are vital. Our specific focus in this paper is on rich types that articulate the properties of tabular operations. We wish to study both their *expressive power* and *diagnostic quality*.

**Inquiry**  There is no "standard library" of table operations. As a result, every paper (and project) is free to use its own (sub)set of operations. This makes artifacts very difficult to compare, and it can be hard to tell whether omitted operations were left out by oversight or because they cannot actually be expressed. Furthermore, virtually no papers discuss the quality of type error feedback.

**Approach**  We combed through several existing languages and libraries to create a "standard library" of table operations. Each entry is accompanied by a detailed specification of its "type," expressed independent of (and hence not constrained by) any type language. We also studied and categorized a corpus of (student) program edits that resulted in table-related errors. We used this to generate a suite of erroneous programs. Finally, we adapted the concept of a datasheet to facilitate comparisons of different implementations.

**Knowledge**  Our benchmark creates a common ground to frame work in this area. Language designers who claim to support typed programming over tables have a clear suite against which to demonstrate their system's expressive power. Our family of errors also gives them a chance to demonstrate the quality of feedback. Researchers who improve one aspect—especially error reporting—without changing the other can demonstrate their improvement, as can those who engage in trade-offs between the two. The net result should be much better science in both expressiveness and diagnostics. We also introduce a datasheet format for presenting this knowledge in a methodical way.

**Grounding**  We have generated our benchmark from real languages, libraries, and programs, as well as personal experience conducting and teaching data science. We have drawn on experience in engineering and, more recently, in data science to generate the datasheet.

**Importance**  Claims about type support for tabular programming are hard to evaluate. However, tabular programming is ubiquitous, and the expressive power of type systems keeps growing. Our benchmark and datasheet can help lead to more orderly science. It also benefits programmers trying to choose a language.

## The Art, Science, and Engineering of Programming

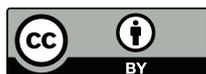

## 1 Motivation

Tables are a widely-used means of communicating information. They suggest a clean and useful visual representation, and they save data-processors from parsing. They are readily understood and created even by children [36]. Thus, it is unsurprising that a large quantity of data—e.g., government data repositories about everything from demographics to voting to income—are provided as tables (often as CSV files). Furthermore, many datasets are provided in siblings of tables such as spreadsheets and relational databases.

In turn, many programming languages support tabular programming. Some, like SQL, are custom languages, but for many programmers, it is convenient (especially when tables are of moderate size, so that performance is less of a concern) to process tables from within whatever language they are using to write a larger application, or with which they are already very comfortable.

Tables are a rich source of typing discipline. Typically, each column of a table is homogeneous, but the columns can themselves vary in type. There can also be a large number of columns. This makes the typing of tables interesting. Is it just Table? That provides no information about values extracted from a table. Is it Table<T>? That implies all data in the table are homogeneous, which is rarely the case. Is Table a constructor with a type per column? Given that tables do not have a fixed number of columns, this requires variable arity for constructors. Columns are usually accessible by name. They are often also ordered. And so on (section 3).

Owing to this richness and complexity, several authors have created sophisticated systems to type tables in higher-order, functional programming languages (section 10). Furthermore, tables offer an opportunity for authors of new, richly-typed languages to showcase what their language can do. We are especially interested in these typing schemes because we are currently designing a typed table library for the Pyret language. If there are known techniques that could meet our needs, we would be delighted to use them.

Unfortunately, there is a significant difficulty in performing scientific comparisons among table types, which we saw first-hand. In Spring 2021, the authors ran a graduate seminar to study the state of the art in rich type systems to support programming with tables.[1] Our focus was on both expressiveness and human-factors—the latter (e.g., error quality) often being the victim of enrichments to the former.

We were left deeply frustrated for two reasons. First, we saw little discussion of human-factors in most of the papers. Second, and even more notably, it proved very difficult to compare the many papers we read. There are many operations over tables, but most papers discussed only a very small, select set of them. Were other operations left out due to pure oversight, because of space, or because they were beyond the power of the type system being proposed? Even when operations were present, they were sometimes in a weaker form than one might wish—leaving the same questions

---

[1] cs.brown.edu/courses/csci2950-x/2021 accessed 2021-08-31

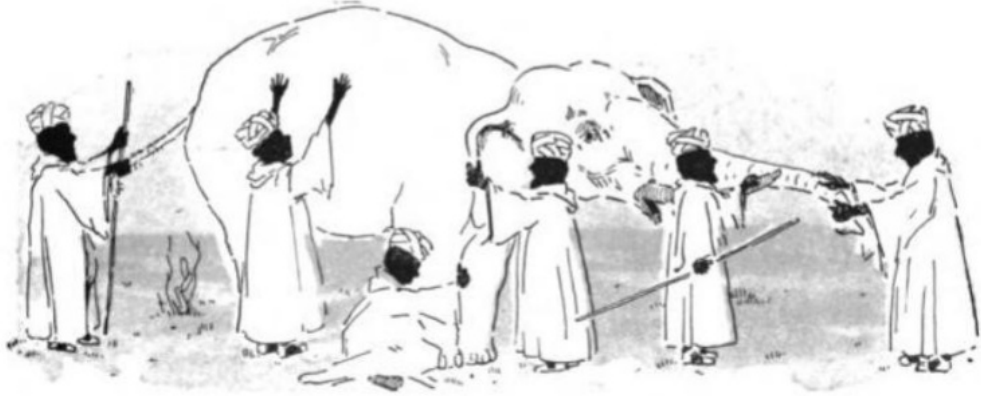

**■ Figure 1** Live View of Type System Designers Approaching Tabular Programming [54]

unanswered. Ultimately, our efforts to summarize the research landscape (section 10) were curbed by the narrow focus of the papers (informally characterized in figure 1).

We want to be clear that we do not blame any of these authors. Typing tables is a large problem. There are many table processing operations to support. Operations have complex behavior, so their type can be expressed at different levels of richness. Error reporting is another challenge. Each paper has advanced the state of the art by imposing constraints and focusing on part of the problem, which is a reasonable way to make progress.

However, we believe it would be healthy for the community to have a fixed reference point to use as a comparison. Some researchers may be able to show that their systems can already capture all these constraints. Others may be spurred on to new research challenges. Authors of new richly-typed languages can validate their advances over prior work. Researchers who make progress on error reporting can use an objective, third-party suite to demonstrate their improvements. Finally, programmers can use this suite as a guide to understand completeness, richness, and diagnostic aid when choosing languages. Thus, we anticipate salutary outcomes for both the research and programming communities.

While the benchmark provides axes of standardization, it does not automatically dictate how results should be reported. Wide variation in reporting is therefore a risk, and would vastly reduce the value of the benchmark. However, because type systems and programming languages are engineering media, ideas from other forms of engineering may help here. We propose adapting the idea of a datasheet to standardize reporting. In this paper, we introduce a first version of such a datasheet (section 8).

## 2 Benchmark Components and Design

This paper's contribution is b2t2, the Brown Benchmark for Table Types. The purpose of the benchmark is to serve as a focal point for research on type systems for tabular programming. To this end, we have collected a set of table operations that is sufficient for many tasks and have annotated these operations with precise constraints for type systems to strive for. In addition to this central Table API, the benchmark has four auxiliary components to validate the end-to-end experience of a proposed type system:

1. *What Is a Table?* (section 3). A definition of tables. The definition gives candidate type systems a common starting point.

2. *Example Tables* (section 4). A set of example datasets. These are mainly used to illustrate the behavior of operations in other components of B2T2, but also concretize some expressiveness challenges.

3. *Table API* (section 5). An API of tabular operations with detailed type constraints. The constraints are intentionally writted in English to set goals for type system designers in a neutral manner.

4. *Example Programs* (section 6). A set of short programs that use the table operations. Each program raises specific typing questions, such as how to iterate over the numeric columns in a table.

5. *Errors* (section 7). A set of buggy programs with explanations that describe the exact bug in the program. The challenge for a type system is to provide accurate, and ideally actionable, feedback.

To encourage comparisons among implementations of B2T2, the benchmark concludes with a datasheet template for reporting purposes:

6. *Type System Datasheet Template* (section 8). A short free-response form that authors can use to describe their implementation of B2T2.

## 2.1 Benchmark Source

B2T2 is implemented as a public GitHub repository:

<div align="center">

github.com/brownplt/B2T2

</div>

Our vision is for the benchmark to serve as a living artifact for the community. First of all, we encourage type system designers to contribute implementations of the benchmark. Second, we invite criticism of the benchmark itself. This initial release may contain implicit assumptions, despite our efforts to weed them out. Similarly, other researchers may propose new elements to enrich the main components—characteristics, example tables, operations, programs, or errors.

## 2.2 Benchmark Origins

We have constructed B2T2 drawing on several major tabular programming frameworks: R (Tidyverse) [19, 66], Python (pandas) [57], Julia [31], LINQ [43], and SQL [17]. We chose these sources because they are widely regarded as quality data-processing tools and are used in numerous domains in industry and elsewhere. Thus, they provide us with a sense of authenticity.

Readers will notice that the frameworks are mostly from "dynamically typed" languages. This choice is intentional. Starting with a typed library might restrict us to operations that are only expressible in those type systems. In dynamic languages, programmers are not hampered by any particular static type system and can more freely express the behaviors that they find convenient.

B2T2 is meant to be a compelling yet moderately-sized challenge for type system designers. Thus, rather than re-create a full library in the same vein as the inspiring frameworks, we have taken steps to curate an API. In particular, we have:

1. Been selective in copying operations, focusing on a sufficient set of operations to reveal challenges for type systems. For example, the common operation nlargest is missing because it can be encoded with two API operations: tsort and head.

2. Limited the set of parameters and options of the ones we do copy. For instance, B2T2 distills the eight cases of the pandas join operation down to one.

3. Minimized the use of overloading. Instead of including one concat operation that can append rows or columns depending on a parameter, B2T2 contains two operations: vcat and hcat. Section 5 explains the minimal overloading that is included.

4. Translated overly-dynamic features. For instance, some of these libraries let programmers pass expressions in the form of strings. Not only does this depend on an eval-like feature in the language, it also leaves ambiguous how scope is handled. We have instead used function parameters, which avoid all these problems.

In addition to the frameworks, B2T2 is inspired by one more, rather different, source of inputs: Pyret, an educational language, as used in the Bootstrap:Data Science [9] curriculum; and the data-centric [37, 53] collegiate curricula. The use of these curricula serves as a check on the sufficiency of our operation choices; it ensures that the Table API can clearly represent fundamental data-processing tasks.[2]

## 2.3 Design Alternatives

Before we present the components of B2T2, we pause to discuss a few significant non-goals of the design:

- B2T2 is meant to improve the *normal design* [63] of type systems. Although (aspects of) the benchmark may be useful in other settings—as a reference point for a new domain-specific language, as a source of example programs and errors, etc.—we fully acknowledge that B2T2 is not an appropriate tool to validate *radical* [63] category-breaking methods of tabular programming. Section 9 discusses this limitation in more detail.

- On a similar note, B2T2 is a benchmark for expressiveness aspects of a type system. It is not concerned with the efficiency of operations. Nor does it include broader approaches to evaluation such as cognitive dimensions [8] and conceptual design [28].

- There are several aspects of the programming experience that B2T2 does not cover. Most of all, we mention the ergonomics of the programming interface. Today, one can program using a variety of tools, from structured editors [58] and block-based programming [40] to dot-driven metaphors [48] and new modalities [2, 47, 49]. It has not been clear how to capture these in a technical manner, and some of these anyway stray quite far afield from our focus on types. Section 7 briefly returns to this issue.

---

[2] Full disclosure: We hope to one day design a richly-typed table library for Pyret.

## 3  What Is a Table?

github.com/brownplt/B2T2/blob/v1.0/WhatIsATable.md

B2T2 begins with a definition of what we consider to be a table, since the term does not have complete agreement. We intentionally do not over-specify the definition to avoid precluding some clever encoding that we have not envisioned. Rather, we list those characteristics that we consider essential and highlight key design choices.

### 3.1  Basic Definitions

- A *table* has two parts: a schema and a rectangular collection of cells.
- A *schema* is an ordered sequence of column names and corresponding sorts.
  - The column names must be distinct (no duplicates).
  - The sorts can vary freely.
- A *header* is a sequence of distinct column names (a schema without sorts).
- A *column name* is a string-like first-class datatype.
- A *sort* describes the kind of data that a cell may contain.

> *Common sorts are numbers and strings; uncommon sorts include images, sequences, and other tables.* [3]

- The collection of cells has $C * R$ members, where:
  - $C$ is the length of the schema;
  - $R$ is an arbitrarily-large number of rows; and
  - each cell has a unique index $(c, r)$ for $0 \leq c < C$ and $0 \leq r < R$.

> *The rectangular arrangement has four important consequences: the rows are ordered, the columns are indexable by schema, all columns contain exactly R cells, and all rows contain exactly C cells.*

- A *row* is an ordered sequence of cells.
- A *cell* is a container for data.
  - Cells may be empty.
  - The data in cells of column $c$ must match the sort of the $c$-th element of the schema.

---

[3] We use the term "sort" to avoid any intuitive constraints that readers might attach to the term "type". All types are sorts. If needed, a sort can be more.

## 3.2 Additional Characteristics

- Empty tables, with no cells, may have zero rows and/or zero columns.
- Tables can be represented in row-major order, in column-major order, or in any other way that supports the basic definitions and the Table API operations. Encodings of tables that use other abstractions (functions, objects) are quite welcome as well.
- For this first version of the benchmark, we assume that tables are immutable. Supporting mutation adds significant but largely orthogonal complications: some systems may need to layer on effect systems [38], and all systems have to manage aliasing for soundness. Furthermore, many rich type systems are built atop purely-functional languages. Since one reason to permit mutation is for efficiency, a topic we are explicitly ignoring, this omission does not create problems elsewhere.
- Column sorts are not first class. Nor do we assume any reflective operations on sorts. Consequently, a program cannot filter the numeric columns from a table by inspecting column sorts and the describe/summary functions of R, pandas, and Julia are inexpressible.
- Finally, we ignore input-output: the benchmark does not stipulate how tables are entered into programs. There may be a variety of mechanisms: typed in verbatim, loaded from a file, inserted using a drag-and-drop interface, and so on. We only expect that there be some way to express table constants like those in section 4.

## 3.3 Typing Tables

Tables have several features that present challenges for conventional type systems, especially because table operations can manipulate aspects of a table. We list these features and their justification below:

- *Columns are heterogeneous.* The column sorts in a table schema allow different kinds of data to sit side-by-side. As a basic example, a table may have numbers in its first column and strings in its second. This property is critical to describe existing datasets, but it does not fit with type systems that require homogeneous collections. Although programmers can create homogeneity by defining an artificial "supertype" that unifies all the actual types contained in a table, this extra step is an imposition that complicates the boundary between datasets and the programming language.
- *Cells may be empty.* Real-world datasets often lack some entries. It is therefore critical that tables can express empty cells. We do not mandate a particular choice, however, because determining how to represent missing values is complex issue that may vary across languages (table 1), and these debates have an even longer history in databases [13, 15, 45].
- *Rows and columns are ordered.* Rows are ordered so they can be referenced by index; we ignore here performance issues such as random access. Columns are ordered so that users can keep salient columns side-by-side to compare them visually. (Think about the times you've reordered the columns of a spreadsheet to put a pair of interesting columns beside each other.) Tables must preserve this order at least in their presentation, whether or not they do in their internal representation.

■ **Table 1** The defining characteristics of our tables and a comparison to related work.

| | Table | R | | pandas | Julia | |
| --- | --- | --- | --- | --- | --- | --- |
| | | Tibble | D.Frame | D.Frame | D.Frame | D.Table[d] |
| Rectangular | ● | ● | ● | ● | ● | ● |
| Ordered | ● | ● | ● | ● | ● | ● |
| Column Names | ● | ● | ● | ● | ● | ● |
| Distinct Columns | ● | ● | | ○ | ● | ● |
| Row Names | | | ● | ● | | |
| Implicit Null | ? | ● | ● | | ● | |

   d Julia DataTables are deprecated as of May 2021.

- *Column names are first-class and manufacturable*. There are programs for which it is eminently useful to compute the names of columns dynamically. The quizScoreFilter program (section 6) is one example. Of course, such programs are difficult to type because column names are more than atomic labels. Names must be first-class values and require at least append and split operations to build new columns and to compare with existing ones.[4] Other useful operations include pattern-matching on column names, constructing names from other data (e.g. strings and numbers), and building sets of names via unions, intersections, and complements.

### 3.4 Design Alternatives

Table 1 presents a more detailed comparison among B2T2 tables, R tibbles (a modern refinement of R data frames [66]), and the data frames found in R, pandas, and Julia. The rows describe notable features. A filled circle indicates the default presence of some feature, a blank space indicates an absence, and an open circle indicates a feature that is configurable but disabled by default. Because B2T2 tables are a specification rather than an implementation, the row for implicit null uses a third symbol ? to mark an unspecified feature.

All designs have *column names* and impose both a *rectangular* shape and *ordered* rows and columns. Designs disagree on the other features: whether columns must be distinct, whether rows have names, and whether there is an implicit notion of null to represent missing data.

- Column names are *distinct* in tables, in tibbles, and in both Julia libraries. The others allow duplicate columns by default and distinguish these columns by position. B2T2 requires distinct columns so that table operations can raise a type error if two columns have the same name. Furthermore, disallowing duplicates may make it easier for SMT-solver-aided type systems to encode schemas.

---

[4] A language can have first-class names that are not manufacturable; e.g., RASCAL [35].

- Data frames in R and pandas come with *row names*. These names can enable both a handy syntax for accessing data and efficient storage strategies. As a case in point, the F# Deedle library indexes time series data by name rather than position and allows nearest-neighbor lookups.[5] B2T2, by contrast, attaches no metadata to rows. Names must be stored in a column. Incidentally, the tidy data [67] method argues against row names.

- Lastly, the designs are split about whether to encode missing data with an implicit null or not. Tibbles, R data frames, and Julia data frames each come with a sentinel value that may be appear in any column and propagates through common operations. Data frames in pandas do not have a uniform treatment of null; certain Python/pandas types have a null value (e.g. float has np.nan), but other types lack an idiomatic default.[6] Julia data tables do not support null, and instead require the use of an Option type. Interestingly these explicit-null data tables were proposed as an enhancement over Julia data frames, but caused enough breaking changes to warrant a separate package.[7] B2T2 leaves this contentious decision to implementors and merely illustrates situations in which missing data can arise.

## 4   B2T2: Example Tables

github.com/brownplt/B2T2/blob/v1.0/ExampleTables.md

The second component of B2T2 is a curated set of tables that highlights the basic challenges in representing tabular data. These tables also serve as concrete examples for other parts of the benchmark.

The example tables have the following characteristics:

- They are intentionally small. This is because some languages (especially core languages) may require verbose encodings. There is value to seeing how a small table looks in any system, to compare the systems' ergonomics.

- They contain values that range over a small, but representative set of sorts: numbers, booleans, strings, sequences, and sub-tables.

- Some tables, like gradebookMissing (figure 2), contain empty cells; we indicate these using blank spaces. Each encoding must determine how to handle these.

- The tables jellyAnon and jellyNamed are designed to support operations that iterate over all columns, selecting just those with boolean values.

- The tables employees and departments are designed for use in join operations.

| name | age | quiz1 | quiz2 | midterm | quiz3 | quiz4 | final |
|------|-----|-------|-------|---------|-------|-------|-------|
| Bob | 12 | 8 | 9 | 77 | 7 | 9 | 87 |
| Alice | 17 | 6 | 8 | 88 | | 7 | 85 |
| Eve | 13 | | 9 | 84 | 8 | 8 | 77 |

■ **Figure 2** Example table gradebookMissing

## 4.1 A Sample Table

Figure 2 illustrates the gradebookMissing example table. This gradebook resembles data that an instructor might keep. Each row corresponds to a student. The first few columns describe the student. The remaining columns contain numeric grades for different assignments. Note that several column names share a common prefix, quiz, which motivates two table-processing tasks: adding a column (to store the results of a new quiz), and dynamically computing column names (quizScoreSelect in section 6). Two cells are empty, perhaps because the student was absent on a quiz day.

This example table, like others in B2T2, shows only a header instead of a full schema. We let implementations choose appropriate sorts for each column. Of course, implementations are free to require schemas on all table literals.

## 4.2 Design Notes

The example tables are intended as small illustrations. This narrow focus means that some potential design goals are not met by this first version of the benchmark:

- The example tables are not intended to reflect principles of good test-case design (e.g., various edge cases), and are not meant to form a sufficient test suite.
- These tables also ignore various considerations that arise from programming languages, mathematics, or specific domains. Such considerations include number representations, statistical spread, and serializability.

## 5 B2T2: Table API

github.com/brownplt/B2T2/blob/v1.0/TableAPI.md

The third and central component of B2T2 is a functional API that supports common data-processing tasks. Each entry in this Table API comes with a conventional sort, a (possibly empty) set of requirements, and a (non-empty) set of guarantees. The challenge for language designers is to express these constraints with types and code.

---

[5] fslab.org/Deedle/series.html accessed 2021-08-31

[6] pandas.pydata.org/pandas-docs/stable/development/roadmap.html#consistent-missing-value-handling accessed 2021-08-31

[7] github.com/JuliaData/DataFrames.jl/issues/1148 accessed 2021-08-31

### 5.1 Design Goals, Characteristics

- As a collection of operations, the goal of the Table API is to express idiomatic tasks—such as those needed by the curricula noted above (section 2.2)—and to highlight issues for type system design.

  - The API is not meant to be a "core" definition that chooses a minimal set of operations. Rather, the focus is to identify the common tasks for which a core definition should provide a foundation.

  - The API is not meant to serve as a full-fledged table library. For example, it omits a handy subTable operation because that behavior can be expressed as a composition of two included operations: selectColumns and selectRows.

- The requirements and guarantees that annotate each API operation are meant to be complete specifications that describe all properties that a type system might enforce. Additionally, the require/guarantee specifications are written in English to avoid bias toward any particular type system.

- The API includes two pivot operations to support the tidy data style of data cleaning [67]. Notably, these operations show the importance of first-class column names. They require the insertion of column names into table cells and the projection of column names from cells into the schema.

- None of the API operations depend on first-class sorts. Type system designers need not support reflection on types to express the Table API.

### 5.2 A Sample API Entry

An API entry presents a name, a conventional sort, pre and post conditions, a prose description, and simple examples. Figure 3 sketches the way that these pieces come together for one entry.

### 5.3 API Format and Conventions

A full description of the API would not make for interesting reading. Therefore, we defer documentation of the full API to the repository link at the top of this section. We do, however, need to explain a few notational conventions in the API that are not intended as constraints on language designers.

1. The Table API splits overloaded operations into separate definitions. The selectRows operation, for example, expects a table and a sequence that describes which rows to extract. Given a sequence of numbers, it selects the corresponding rows. Given a sequence that contains one boolean per row, it selects using the indices of true values in the sequence. Thus, the API has two definitions:

   ```
   (overload 1/2) selectRows :: t1:Table * ns:Seq<Number> -> t2:Table
     ....
   (overload 2/2) selectRows :: t1:Table * bs:Seq<Boolean> -> t2:Table
     ....
   ```

---

addColumn :: t1:Table * c:ColName * vs:Seq<Value> -> t2:Table

    *Where t1, c, vs, and t2 name the respective parts of the sort.*

**Constraints**

Requires:

- c is not in header(t1)
    - *i.e., c must be a fresh column name*
- length(vs) is equal to nrows(t1)
    - *i.e., the sequence of values must have exactly one element per row*

Ensures:

- header(t2) is equal to concat(header(t1), [c])
- for all c' in header(t1), schema(t2)[c'] is equal to schema(t1)[c']
- schema(t2)[c] is the sort of elements of vs
- nrows(t2) is equal to nrows(t1)

**Description**

Consumes a column name and a Seq of values and produces a new Table with the columns of the input Table followed by a column with the given name and values. Note that the length of vs must equal the length of the Table.

```
> hairColor = ["brown", "red", "blonde"]
> addColumn(students, "hair-color", hairColor)
```

| name | age | favorite color | hair-color |
|------|-----|----------------|------------|
| Bob | 12 | blue | brown |
| Alice | 17 | green | red |
| Eve | 13 | red | blonde |

■ **Figure 3** Example API entry

Language designers are welcome to handle overloaded operations in any way makes sense in their language: overloading, distinct operations with related names, subclasses, or something else.

2. If an operation can fail, then its result sort is Error<T> for some sort T. Implementors will need to express error terms in an idiomatic manner (perhaps with a tagged message or an integer flag) and may need to adapt the sorts of such operations.

3. The API uses higher-order functions and other forms of abstraction for convenience. Language designers do not need to support exactly these abstractions as long as they can express a similar behavior, perhaps through inlining. For example, buildColumn expects a function that creates a new value from a row, applies this function to each row, and collects a new table column.

  buildColumn :: t1:Table * c:ColName * f:(r:Row -> v:Value) -> t2:Table

A first-order language might underapproximate this behavior by proscribing a pattern that users can follow after they have defined an f function.

    The orderBy operation presents a more difficult example because conventional types give only a vague impression of its behavior. This operation, which is a com-

bination of the lazy `OrderBy` and `ThenBy` methods of LINQ Enumerables [16], uses a sequence of pairs of functions to lexicographically sort a table. Each pair consists of a getKey function and a compare function such that the result sort of getKey matches the input sort of the compare at hand. Different pairs can employ different sorts; e.g., the first compare may expect numbers while the second compare expects strings.

```
orderBy :: t1:Table
            * Seq<Exists K . getKey:(r:Row -> k:K) * compare:(k1:K * k2:K -> Boolean)>
            -> t2:Table
```

One indirect way to express this behavior is to ask for a single getKey function that returns a tuple and a compare function that lexicographically compares the tuples.

4. We acknowledge that the API is written with rather flexible structural typing in mind. Consider the sort for `buildColumn` above; it assumes that the language can collect a sequence of values into a column and then append that column to widen a table. A language might not support such operations in a fully-generic manner (perhaps to enable Hindley–Milner inference), in which case it is acceptable for a language's `buildColumn` to ask for a function that computes an entire row. Other operations may require analogous details to explain how pieces fit together. For instance, the sequence argument to orderBy may be easier to express as a heterogeneous tuple.

## 5.4 Conformance

The overall purpose of the Table API is to set goals for language designers and to enable comparisons among implementation efforts. An explicit non-goal is to constrain the form of tabular languages and type systems. To help clarify this non-goal, we list several points that the API does not mandate.

- The Table API is not a core definition and does not require any particular set of primitives. An implementation may begin with its own core set of primitives and use those to express API operations.

- Similarly, an implementation need not express each API operation as a standalone function. Other ideas include: syntactic sugar (macros), methods, and compositions of other operations. Depending on the choice, an implementation may change the presentation of API entries to match. For example, methods may require different signatures with one fewer parameter.

- Implementations are welcome to choose entirely different *names* for API operations. Though, to aid comparison, it would be helpful to accompany such changes with a map to the names in B2T2.

- Implementations are also welcome to clarify the *sorts* for operations. The unassuming Table sort almost certainly requires parameters. The generic Seq sort may need to be specialized, perhaps to vectors in some cases and to tuples in others. Languages that track nullable values may need more-precise signatures to clarify which arguments can and cannot be null.

- Implementations need not encode all these properties in the pre and post conditions. Our only requirement is that implementations give an explanation of what is and is not (or, cannot be) expressed to enable comparisons.

## 6  B2T2: Example Programs

github.com/brownplt/B2T2/blob/v1.0/ExamplePrograms.md

B2T2 contains a set of example programs to test how well an implementation of the Table API supports the development of new typed code. Of particular interest is code that uses type system features not found in the API. Each example has two parts: a problem statement and a reference implementation. We list the problem statements below, with emphasis on the type system challenges that each one presents.

**dotProduct**  This function computes the dot product of two columns in a table (otherwise known as SUMPRODUCT in Excel). A type system should ensure that the columns are in the table and that the sorts of these columns describe numbers. A unit checker might also ensure that the numbers have compatible units.

**sampleRows**  This function selects a random sample of a table's rows. Versions of it are found in many popular table libraries. We choose to include it here because it is effectively stateful, which may make it unwieldy or impossible to express in some languages or type systems. Furthermore, randomness is a particular kind of state that is glossed over by some type systems and not by others. The centrality of randomness and sampling in statistical computation makes it important for users of a programming medium to know how randomness, specifically, will be handled.

**pHackingHomogeneous**  This function illustrates the principle of *p*-hacking using a jellybean dataset inspired by an XKCD cartoon [71]. All columns in the dataset are boolean-valued.

**pHackingHeterogeneous**  This illustrates the same *p*-hacking principle, but against an initial dataset where *not* all columns are booleans. Thus, a system that can type the previous example cannot necessarily type this one.

**quizScoreFilter**  This example describes a task that many instructors perform at the end of a course: compute the average quiz score for each student in a gradebook. The gradebook contains a mix of numeric and non-numeric fields, and the numbers denote both quiz scores and exam scores. To find the quizzes, this example iterates through all column names and filters the ones that begin with quiz.

**quizScoreSelect**  This example also computes the average quiz scores for a gradebook. It does so by appending the column name quiz to a few integer suffixes and selecting these computed columns from the gradebook.

**groupByRetentive**  This example categorizes rows of an input table into groups based on the values present in a key column. The output table includes the key column. The Table API describes a similar operation; we include it in both places to check that user-defined functions can express detailed type constraints.

```
> pHacking =
    function(t):
      colAcne = getColumn(t, "get acne")
      jellyAnon = dropColumns(t, ["get acne"])
      for c in header(jellyAnon):
        colJB = getColumn(t, c)
        p = fisherTest(colAcne, colJB)
        if p < 0.05:
          println("We found a link between " ++ c ++ " jelly beans and acne (p < 0.05).")
        end
      end
  > pHacking(jellyAnon)
```

■ **Figure 4**  Example program pHackingHomogeneous

**groupBySubtractive**  This example categorizes rows of an input table into groups based on the values present in a key column. The output table does not include the key column. Like the previous example, its purpose is to test that user-defined code is no less expressive than API code.

### 6.1  A Sample Program

Figure 4 presents the example code for the pHackingHomogeneous example. It defines a function named pHacking and invokes this function on one of the B2T2 example tables. The function body sketches an implementation using API operations, general-purpose syntax (e.g. loops and conditionals), and one statistical function (fisherTest):

### 6.2  Conformance

For language implementors, the ground rules for the Table API apply to the example programs. It is not necessary to express each example strictly as a function, nor to follow the code line-by-line. Furthermore, we can imagine that some programs cannot be expressed as written, with a (functional) abstraction, but can be typed if parts of the code are inlined. It may also be necessary to rewrite the programs to reveal some information to the type system. An implementation should document such variances.

## 7  B2T2: Errors

github.com/brownplt/B2T2/blob/v1.0/Errors.md

For expressive type systems, effective error reporting can be a major challenge. Thus the final component of B2T2 is a suite of erroneous programs with ground-truth explanations. Each program raises two main questions:

1. Does the type system detect the error?

| name | age | favorite color |
|------|-----|----------------|
| String | Number | String |
| I2 | Bob | blue |
| I7 | Alice | green |
| I3 | Eve | red |

The rows disagree with the schema on the ordering of the first two columns.

■ **Figure 5**  Malformed table constant swappedColumns

2.  If so, how understandable is its explanation?

There is an implicit third question; namely, is the program even expressible? We address this point below (section 7.2) as part of a larger discussion about how to compare error feedback across languages.

All the examples are based on *actual* errors from a log of student programs (submitted anonymously and voluntarily) in an introductory course at Brown University. Our presentation distills the student programs to eliminate unnecessary, confusing, or personally identifying context, rename variables, refer to our sample tables, etc., while leaving the essence of the problem unchanged. Every entry implictly makes assumptions about a student's intent; in some cases, this is difficult to discern just from the erroneous program. In all cases, we therefore studied the edits that the student subsequently made (which typically ended in a corrected version that ran properly) to understand what they meant to write instead of the erroneous program.

Several examples contain an error due to a malformed table constant. Figure 5 presents one example; the data in this table does not match sorts in its schema. These "obvious" mistakes are nevertheless common, and their inclusion gives table-aware languages a chance to showcase their helpful feedback. By contrast, languages that encode tables with a desugaring may struggle to explain such errors in terms of the surface syntax. The benchmark includes several constant errors because we anticipate that a language may give better feedback to some than to others. For example, the output for an empty cell could be very different from that for an empty row. The latter may give a very simple error whereas the former produces an indecipherable one (due to desugaring, etc.). Or it could be the other way around, with a smart error for the empty cell because of contextual heuristics, and an ugly error for the missing row because there is no context.

### 7.1  A Sample Error

Each error entry contains five parts: the names of any example tables (section 3) that the program refers to, the program's intent, the buggy program, an explanation of why the code is erroneous, and a corrected version of the program. Figure 6 presents an error involving two column names and a boolean operator.

**blackAndWhite**

**Context**

jellyAnon

**Task**

The programmer was asked to build a column that indicates whether "a participant consumed black jelly beans and white ones."

**A Buggy Program**

```
> eatBlackAndWhite =
    function(r):
        r["black and white"] == true
    end
 > buildColumn(jellyAnon, "eat black and white", eatBlackAndWhite)
```

**What Is the Bug?**

The logical and appears at a wrong place. The task is asking the programmer to write r["black"] and r["white"], but the buggy program accesses the invalid column "black and white" instead.

**A Corrected Program**

```
> eatBlackAndWhite =
    function(r):
        r["black"] and r["white"]
    end
 > buildColumn(jellyAnon, "eat black and white", eatBlackAndWhite)
```

■ **Figure 6** Example error entry

### 7.2 Towards Error Evaluation Criteria

Analyzing the quality of feedback is not a science, and requires some interpretation. Several recent surveys have analyzed the quality of error messages, and thus offer suggestions of techniques [6, 7, 60]. Human factors research on warning label design is also relevant [69]. Our own prior work develops a robust rubric for analyzing error quality based on subsequent programmer actions [41], and also develops a "static" criterion that applies precision and recall to evaluating the quality of messages at design-time [70]. All these ingredients may prove useful to compare error feedback.

One important aspect of benchmarking errors is that the language may have means to preclude their construction entirely. In a traditional, textual language, one can write virtually any text string and submit it for analysis. In contrast, structured editors and block-based editors have a critical property: it is impossible to construct a syntactically ill-formed program. We will call these *preventative* programming media. Types can be considered an extension of this: they are a context-sensitive well-formedness check, and can thus be incorporated into a preventative editor. The burden then shifts to a qualitatively different kind of phenomenon: from explaining an error that *has occurred* (which can reference a concrete program) to explaining why a certain program *cannot be built* (which pertains to a set of programs that, by definition, cannot exist).

We note that there can be a subtle interaction between preventative media and rich type systems. Block languages, for instance, work well because there are obvious visual cues showing why one block cannot be placed inside another. However, when type errors become more subtle—and context-sensitive—preventative methods have the potential to baffle programmers much more than an after-the-fact error report might [39, 51, 64]. Therefore, this domain presents an interesting case-study in the creation of richly-typed preventative programming interfaces.

### 7.3 Conformance

Unlike the Table API, which permits some freedom of implementation, these error benchmarks are sensitive to small changes. For example, in a sophisticated type system, inlining code can significantly change the detection and, even more so, the reporting of an error. Thus, any deviations must be carefully documented and justified.

### 8   Type System Datasheet Template

github.com/brownplt/B2T2/blob/v1.0/Datasheet.md

Recently, a group of influential data scientists put forward the notion of *datasheets* for datasets [24]. Those authors are directly inspired by a tradition in engineering:

> In the electronics industry, every component, no matter how simple or complex, is accompanied with a datasheet describing its operating characteristics, test results, recommended usage, and other information.

Datasheets enable engineers to quickly compare similar components and choose ones fit for purpose. Noting the importance of datasets and their potential for misuse, the authors present an analogous notion of datasheets for datasets.

Programming languages and type systems would benefit from similar documentation. The goal of such a datasheet is not to preclude innovation or hide novelty or virtues; rather, *on aspects that can be compared*, a datasheet provides a quick, standard way to determine which component can fit a use. Put differently, the benchmark is an attempt to systematize the "input" and the datasheet helps systematically summarize the "output" (i.e., the reporting of the system).

To this end, we accompany B2T2 with a datasheet template for tabular type systems. The authors of a language that implements B2T2 can fill out the datasheet to help would-be readers quickly understand the new language.

### 9   How Not to Use B2T2

As mentioned above (section 2.3), B2T2 is a tool for the normal design of tabular type systems. We developed the benchmark in response to a lack of focus among related works—there are many type systems that support tabular programming in a conventional language, and yet these designs are extremely difficult to evaluate (sec-tion 10). B2T2 sets a common goal for these type system designs. No matter how many constraints a particular type system can express, implementing B2T2 relates the system to the practice of tabular programming as found in widely-used frameworks.

B2T2 is not, however, a suitable goal for all typed tabular languages. Unconventional programming media may struggle to express aspects of the benchmark. For instance, one can imagine a virtual reality-based environment in which users can directly manipulate tables in space; such an environment might have an awkward time with the Table API even if the environment accommodates the example programs. The Subtext language [21], as well as Pyret, synthesize types from example. These media might encode useful constraints without anything resembling traditional code or types. For such media, it makes sense to ignore parts of B2T2. Therefore, B2T2 should never be used to criticize such designs, which may need very different evaluation methods (such as cognitive dimensions [8] and concepts [28]).

In short, B2T2 is a structured and technical evaluation criterion. Although there are many "normal" type systems to which it applies, there may be "slightly abnormal" designs to which it only partially applies and "radical" designs that are out of scope. These radical designs are nevertheless vital for progress in programming media.

## 10    Related Work

**Programming with Tables**    As noted above (section 2), B2T2 is directly inspired by tabular programming frameworks. The frameworks include R Tidyverse [19, 59], Python pandas [42, 57], Julia [31, 32], LINQ [43], and SQL [17, 18]. Each provides a toolkit for comprehending and manipulating tables. The B2T2 Table API selects vetted operations from these sources.

The Bootstrap:Data Science [9] and data-centric computing [37, 53] curricula provide critical validation. First, the teaching materials present basic data science tasks that should be expressible. Second, learners that have taken these courses graciously supplied the logs that we used to find erroneous programs (section 7).

**Types for Tables**    There is a huge amount of prior work on type systems that support tabular programming in some form or another. Because these works pursue different goals and present examples that vary widely in complexity, a direct comparison is difficult. We hope that B2T2 enables apples-to-apples comparisons in the future. To a first approximation, however, the research targets five application areas:

- *Records and Variants*    Any type system that supports polymorphic records can support a kind of tabular programming, e.g. [23, 27, 44, 50, 65], though the details depend on the allowed operations on records. Historically, these systems focus on decidable type inference and disallow first-class labels.

- *Relational Algebra*    The authors of LINQ claim that relational algebra is enough to support programming with a variety of data formats, including tables [43]. Several other languages follow this maxim, including Ferry [26], SML# [10, 46], and Ur [12].

- *Array-Oriented Programming*  Remora is a typed variant of the J programming language [30, 52]. The multi-dimensional array operations in Remora can likely be specialized to tabular programming. Qube [61] and FISh [29] might be repurposed in a similar manner.

- *Data Exploration*  The Gamma (thegamma.net) is an innovative language for data exploration [48]. Programmers import a dataset as an object and type a dot (.) after the dataset name to see a list of analytical operations. Applying an operation computes a new type for the result using a *pivot* type provider [56], which is enabled by an untyped relational algebra engine. The dot-driven programming model is compelling, and we are curious to learn the extent to which it can accommodate other tabular programming idioms.

- *Fancy Types*  The designers of advanced type systems occasionally use tabular programming as an application area to demonstrate expressiveness. Examples include the constant-propagating types in CompRDL [33] and the refinement types of Liquid Haskell [62]. Along these lines, we conjecture that the TypeScript keyof operator [20] can support much of the B2T2 Table API.

**Spreadsheet Programming**  Spreadsheets are not tables, but type systems that detect spreadsheet errors might be useful to detect errors in tables [3, 11, 68]. Tabular programming might also benefit from incorporating some of the more-structured elements of spreadsheet programming, such as reactive equations. These would, however, significantly complicate the APIs.

**Language Design Benchmarks**  We are aware of a few other benchmarks for aspects of language design. One is related and complementary to B2T2: an in-progress expressiveness benchmark [14] that compares several languages and styles of programming on table-processing tasks. The representation design benchmarks [72] identify static representation problems for visual programming languages. POPLMark [5] and POPLMark reloaded [1] present problems for proof assistants. Berkeley Motifs [4] set goals for parallel programming models and architectures. The LWC benchmarks [22] address various aspects of language workbenches, from notation to code reuse. Finally, 7GUIs [34] presents seven tasks for GUI toolkits to express succinctly.

## 11  Conclusion

Ultimately, our goal is to improve the practice of tabular programming via static typing. Rich types have the potential to help all kinds of users; they can offer documentation for learners, performance hints for experts, and feedback for everyone in between. The demand for such tooling is high, and we expect it to grow in the coming years.

B2T2 is a first step toward this long-term goal, designed to focus our own design efforts and to promote scientific discussions with other research teams. It was born of a frustration: trying to reconcile a large number of different papers that each described exciting advances but that were mutually incomparable. We believe the components of the benchmark cover the basics of a quality tabular language. First, we offer a standard definition that conforms well to real-world tables. Second, the example tables concretize the definition. Third, the Table API provides a curated set of operations to strive for. Fourth, the example programs validate the API and illustrate additional typing issues. Fifth, the error illustrations draw attention to diagnostics. Finally, the datasheet template brings these components into a technical summary that is designed to encourage comparisons.

Benchmarks are normative, however, and we acknowledge the threat posed by Goodhart's Law [25]. In brief, the trouble is that "when a measure becomes a target, it ceases to be a good measure" [55]. That said, two points about B2T2 help to lessen its potential of becoming a disconnected measure:

1. We begin with a fairly useful set of operations that represent a foundation for standard table processing. If a language supported only these operations, that would still provide a comfortable programming basis for many situations, especially when general-purpose programming tools are also available.

2. We expect that language designers (and data scientists) will be happy to update our benchmark with examples we have missed, especially—in the spirit of friendly competition—those they support well, or—in the spirit of scientific honesty—those they support poorly.

In sum, we are optimistic that B2T2 will help attract programming language expertise to the central issues for typed tables. Nevertheless, the cautions of section 9 are critical and we would be disappointed if this work were used to squelch innovation.

### Benchmark Links

- GitHub release: github.com/brownplt/B2T2/releases/tag/v1.0
- Zenodo archive: doi.org/10.5281/zenodo.5507463
- Extended paper: cs.brown.edu/research/plt/dl/prog2022-b2t2

**Acknowledgements**   We thank: the attendees of our Spring 2021 seminar for thoughtful discussions; Ben Lerner for suggesting the dotProduct example program; Tomas Petricek for broadening our design discussions; the other reviewers for thoughtful feedback; Titus Barik and Brett Becker for references related to error messages; Margaret Burnett, Will Crichton, Paul Khuong, and Cameron Yick for references to language design benchmarks; Ranjit Jhala for help with Liquid Haskell as we prototyped B2T2; and Matthias Felleisen for early criticism that shaped the direction of this work.

This work was partly supported by the US National Science Foundation. This research was also developed with funding from the Defense Advanced Research Projects Agency (DARPA) and the Air Force Research Laboratory (AFRL). The views, opinions and/or findings expressed are those of the author and should not be interpreted as representing the official views or policies of the Department of Defense or the U.S. Government. Greenman received support from NSF grant 2030859 to the CRA for the CIFellows project.

## A  Datasheet

### A.1  Reference

**Q.** What is the URL of the version of the benchmark being used?

**Q.** On what date was this version of the datasheet last updated?

**Q.** If you are not using the latest benchmark available on that date, please explain why not.

### A.2  Example Tables

**Q.** Do tables express heterogeneous data, or must data be homogenized?

**Q.** Do tables capture missing data and, if so, how?

**Q.** Are mutable tables supported? Are there any limitations?

*You may reference, instead of duplicating, the responses to the above questions in answering those below:*

**Q.** Which tables are inexpressible? Why?

**Q.** Which tables are only partially expressible? Why, and what's missing?

**Q.** Which tables' expressibility is unknown? Why?

**Q.** Which tables can be expressed more precisely than in the benchmark? How?

**Q.** How direct is the mapping from the tables in the benchmark to representations in your system? How complex is the encoding?

### A.3  TableAPI

**Q.** Are there consistent changes made to the way the operations are represented?

**Q.** Which operations are entirely inexpressible? Why?

**Q.** Which operations are only partially expressible? Why, and what's missing?

**Q.** Which operations' expressibility is unknown? Why?

**Q.** Which operations can be expressed more precisely than in the benchmark? How?

### A.4  Example Programs

**Q.** Which examples are inexpressible? Why?

**Q.** Which examples' expressibility is unknown? Why?

**Q.** Which examples, or aspects thereof, can be expressed especially precisely? How?

**Q.** How direct is the mapping from the pseudocode in the benchmark to representations in your system? How complex is the encoding?

## A.5 Errors

*There are (at least) two parts to errors: representing the source program that causes the error, and generating output that explains it. The term "error situation" refers to a representation of the cause of the error in the program source.*

*For each error situation it may be that the language:*

- *isn't expressive enough to capture it*
- *can at least partially express the situation*
- *prevents the program from being constructed*

*Expressiveness, in turn, can be for multiple artifacts:*

- *the buggy versions of the programs*
- *the correct variants of the programs*
- *the type system's representation of the constraints*
- *the type system's reporting of the violation*

**Q.** Which error situations are known to be inexpressible? Why?

**Q.** Which error situations are only partially expressible? Why, and what's missing?

**Q.** Which error situations' expressibility is unknown? Why?

**Q.** Which error situations can be expressed more precisely than in the benchmark? How?

**Q.** Which error situations are prevented from being constructed? How?

**Q.** For each error situation that is at least partially expressible, what is the quality of feedback to the programmer?

**Q.** For each error situation that is prevented from being constructed, what is the quality of feedback to the programmer?

## About the authors

**Kuang-Chen Lu** (LuKuangchen1024@gmail.com) is a PhD student at Brown University.

**Ben Greenman** (benjamin.l.greenman@gmail.com) is a PLT member and a postdoc at Brown University.

**Shriram Krishnamurthi** (shriram@brown.edu) is the Vice President of Programming Languages (no, not really) at Brown University.